\pdfoutput=1
\documentclass[journal,10pt]{IEEEtran}

\usepackage[ruled,linesnumbered]{algorithm2e}
\usepackage{graphicx, subfig}
\usepackage{hyperref}
\usepackage{array}
\usepackage{balance}
\usepackage{amsmath}
\usepackage{booktabs}
\usepackage [capitalize] {cleveref}

\begin{document}

\title{Effective Intrusion Detection in Highly Imbalanced IoT Networks with Lightweight S2CGAN-IDS}

\author{Caihong Wang, Du Xu, Zonghang Li, Dusit Niyato,~\IEEEmembership{Fellow IEEE}, \thanks{Caihong Wang, Du Xu, and Zonghang Li are with School of Information and Communication Engineering, University of Electronic Science and Technology of China, Chengdu, China.

Dusit Niyato is with School of Computer Science and Engineering, Nanyang Technological University, Singapore.

The corresponding author: Du Xu. (Email: xudu@uestc.edu.cn)

This work was partially supported by the National Natural Science Foundation of China (62171085).}}


\maketitle

\begin{abstract}
 Since the advent of the Internet of Things (IoT), exchanging vast amounts of information has increased the number of security threats in networks. As a result, intrusion detection based on deep learning (DL) has been developed to achieve  high throughput and high precision. Unlike general deep learning-based scenarios, IoT networks contain benign traffic far more than abnormal traffic, with some rare attacks. However, most existing studies have been focused on sacrificing the detection rate of the majority class in order to improve the detection rate of the minority class in class-imbalanced IoT networks. Although this way can reduce the false negative rate of minority classes, it both wastes resources and reduces the credibility of the intrusion detection systems. To address this issue, we propose a lightweight framework named S2CGAN-IDS. The proposed framework leverages the distribution characteristics of network traffic to expand the number of minority categories in both data space and feature space, resulting in a substantial increase in the detection rate of minority categories while simultaneously ensuring the detection precision of majority categories. To reduce the impact of sparsity on the experiments, the CICIDS2017 numeric dataset is utilized to demonstrate the effectiveness of the proposed method. The experimental results indicate that our proposed approach outperforms the superior method in both Precision and Recall, particularly with a 10.2\% improvement in the F1-score.
\end{abstract}

\begin{IEEEkeywords}
Deep learning, class imbalance, intrusion detection, generative adversarial networks, internet of things
\end{IEEEkeywords}

\IEEEpeerreviewmaketitle

\section{Introduction}
\label{introduce}
The emergence of the 5G era has brought new challenges to cybersecurity due to the proliferation of the Internet of Things (IoT). IoT devices are known to harbor a significant amount of private information and are often secured with simple encryption. As a result, a considerable number of these devices may be rendered as zombie hosts or utilized as mining tools, with some users even falling prey to cyber extortionists.

Intrusion detection systems (IDS) \cite{liao2013intrusion} constitute a pivotal component of firewalls and can detect viruses before they reach IoT devices. As such, IDS has become an indispensable preventive measure for ensuring the security of IoT networks. Conventional intrusion detection technologies heavily rely on manually crafted rules and signatures. The creation and upkeep of these rules and signatures require significant time and labor. However, the contemporary proliferation of traffic and the rising prevalence of attacks stemming from the IoT have rendered these traditional methods relatively ineffective.

To overcome the limitations of traditional intrusion detection methods, deep learning (DL) has surfaced as a promising approach. DL algorithms, including deep belief networks (DBN) \cite{mohamed2011acoustic}, convolutional neural networks (CNN) \cite{girshick2015fast}, and recurrent neural networks (RNN) \cite{schuster1997bidirectional}, can automatically learn complex patterns and anomalies from raw network traffic. This enables more accurate automatic detection of potential threats. Additionally, DL algorithms can effectively leverage massive network traffic to identify potential attacks and adapt to changing attack patterns, thereby significantly enhancing the detection accuracy of the IDS.

DL algorithm has exhibited high accuracy in detecting network attacks \cite{lecun2015deep}. However, to operate effectively, DL models require an adequate number of training examples. In comparison to the abundance of benign network traffic examples, certain attack categories are scarce in number. Consequently, DL-based intrusion detection models encounter the challenge of a high false-negative rate \cite{aminanto2016deep,liu2019machine,ahmad2021network}.

In the realm of class imbalance, researchers have endeavored to optimize the efficiency of deep learning techniques, including data-level approaches \cite{liu2022intrusion,zhang2019deep}, algorithm-level strategies \cite{khan2017cost,zhang2018cost}, integrated learning \cite{dhote2020hybrid,aljawarneh2018anomaly}, transfer learning \cite{wu2019transfer,singla2019overcoming}, and evaluation metrics \cite{zhang2020novel}, with the primary aim of mitigating the false-negative rate of intrusion detection systems (IDS). However, this objective often comes at the expense of precision for majority classes, while improving the detection rate of minority attacks. Therefore, these methods may ultimately not only compromise the reliability of the system but also waste resources.

The motivation of this paper is to enhance the detection rate of minority categories in IoT networks while minimizing the impact on the detection rate of majority categories. By focusing on the distinctive characteristics of attack frequency, we try to pay more attention to the extremely rare attacks and foster the advancement and innovation of this field from different angles.

In response to the aforementioned concerns, we present a proficient and lightweight S2CGAN-IDS framework that leverages the distribution characteristics of traffic categories within IoT networks. Our framework extends the original imbalanced training data by considering two distinct perspectives: data space and feature space. This approach aims to enhance the detection rate of underrepresented categories while maintaining satisfactory detection rates for the majority classes.

The main contributions of this paper are summarized as follows:
\begin{enumerate}
\itemsep=0pt
\item We have devised a lightweight S2CGAN-IDS framework from a data-oriented perspective to address the issue of class imbalance. This framework aims to improve the detection rate of the underrepresented minority class while maintaining accuracy for the majority class.
\item This paper presents an innovative feature extraction method that combines Siamese networks and autoencoders to preserve class differences and significantly accelerate the convergence speed of the adversarial generative network.
\item This paper presents a novel data augmentation technique, SCGAN, for categories exhibiting similar distribution profiles. The proposed approach, which combines Siamese networks and autoencoders, accelerates the convergence rate significantly.
\item This paper introduces a highly efficient data synthesis approach named synthetic k neighbors (SKN) that utilizes feature space-based methods to generate samples for categories that are extremely rare.
\end{enumerate} 

\section{Motivation}
\label{motivation}
The class imbalance problem in IoT scenarios is of paramount importance in ensuring IoT security \cite{le2022xgboost}. This problem stems from several key factors, including the extensive deployment of devices, the wide variety of malicious behaviors, the limited resources of IoT devices, and the heightened sensitivity of security requirements. These factors collectively contribute to the scarcity of malicious behavior data in IoT scenarios, making accurate detection of such behaviors an urgent necessity \cite{leevy2021mitigating}. Consequently, the effective resolution of the class imbalance problem holds significant significance in upholding IoT security.

After an extensive literature search, NSLKDD, UNSW-NB15, and CICIDS2017 have emerged as the predominant datasets utilized in this field over the past two decades. An evaluation of attack frequency across these datasets reveals a distinct gradient shift. Notably, CICIDS2017 exhibits a conspicuous step-like upgrade while possessing the most recent and sparsest characteristics, aligning it more closely with the traffic observed in real IoT network environments \cite{stiawan2022improvement}. Consequently, CICIDS2017 has been chosen for subsequent analysis and experimentation.
\begin{table}[!h]
  \centering
  \caption{Some details of commonly used classic datasets.}
  \begin{tabular}{p{1.6cm}p{0.8cm}p{2.1cm}p{0.8cm}p{1.2cm}}
    \toprule
    \textbf{Dataset} & \textbf{Year} & \textbf{Characteristics} & \textbf{Sparsity} & \textbf{Frequency} \\
    \midrule
    NSLKDD & 1999 & Network-based, real-world traffic, KDD Cup 1999 & Medium & Uniform         \\
    UNSW-NB15 & 2015 & Network-based, real-world traffic, contains synthetic and real data& High & Gradual          \\
    CICIDS2017 & 2017 & Network-based, real-world traffic, contains IoT and normal traffic & Low & Stepped             \\
    \bottomrule
  \end{tabular}
  \label{tab:ddatasets}
\end{table}

\begin{figure}[!ht]
    \centering
    \includegraphics[scale=0.15]{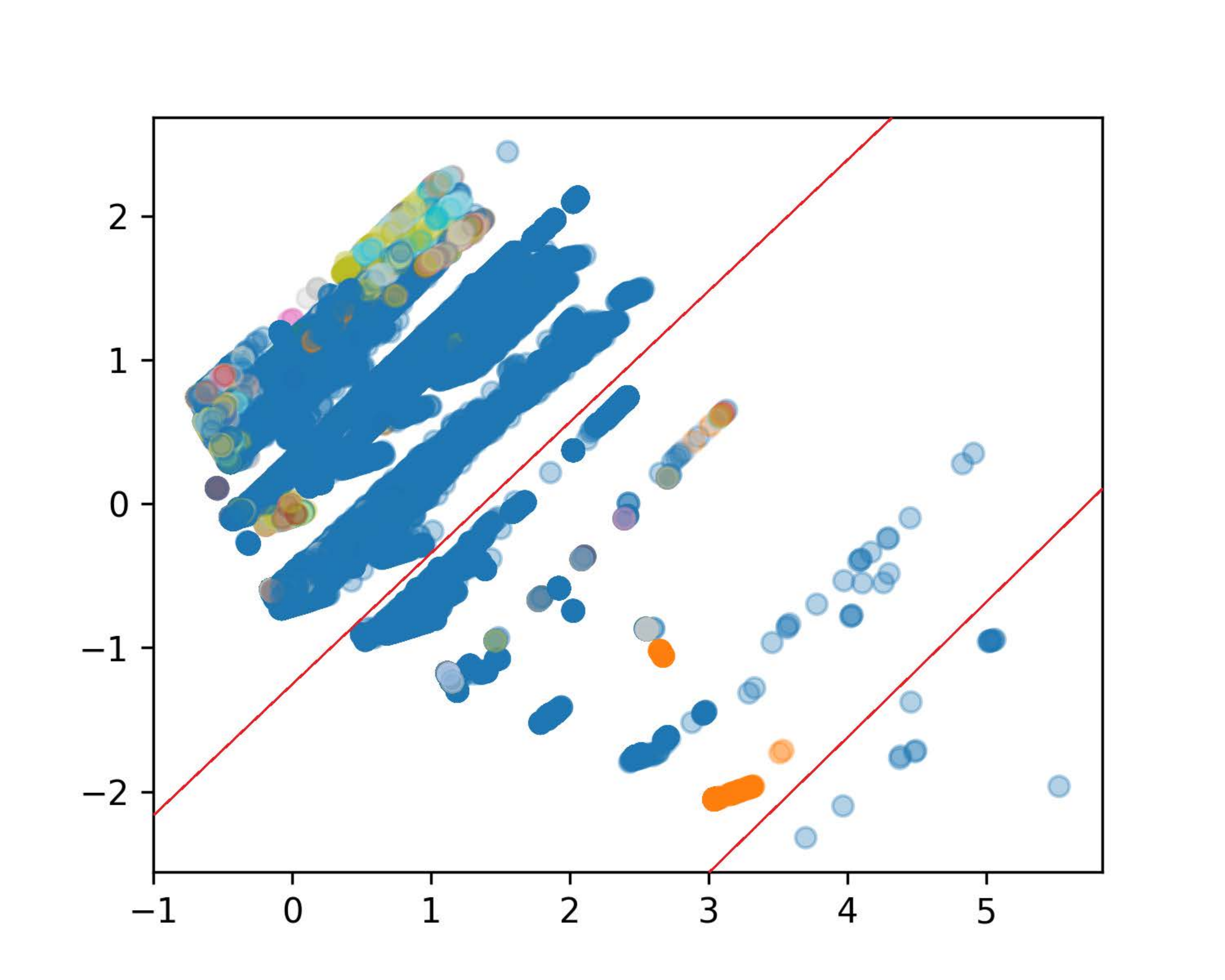}
    \caption{The scatter of CICIDS2017 dataset.}
    \label{fig:scatter}
\end{figure}

Based on the analysis mentioned above, the fundamental issue that must be addressed by an effective IoT network intrusion detection model is enhancing the detection rate of the minority categories while maintaining the accuracy of the majority categories. To tackle this problem, we conducted Principal Component Analysis (PCA) on a widely used intrusion detection dataset and generated a scatterplot based on the resulting PCA data (\cref{fig:scatter}).

Analysis of \cref{fig:scatter} reveals that categories located in the upper-left region of the scatterplot possess an ample number of samples and exhibit a complete distribution. Conversely, the categories in the middle region are relatively scarce, and the distribution outline is rather rough. Finally, categories situated in the lower-right region contain only a few, scattered data points.

Based on the observed characteristics in the scatter, we calculate the imbalance ratio ($I\!R_i$) for each class by $n_{max} / {n_i}$, where $n_i$ represents the number of the $type\text{-}i$ attack, and $n_{max}$ represents the number of normal samples. Remarkably, our calculations revealed a distinct step-wise distribution pattern in the $I\!R_i$ values, which aligned with the visual representation depicted in the scatter.
\begin{table}[!h]
\centering
\caption{The labels, $I\!R_i$ values and quantity levels of the CICIDS2017 dataset.}
\setlength{\tabcolsep}{1.5mm}
\begin{tabular}{ | m{3cm} |m{1.2cm} |m{1.2cm}| m{0.8cm} | } 
    \hline 
\textbf{Subclass} & \textbf{Label} & \textbf{$I\!R_i$} & \textbf{Level} \\
    \hline
    BEINGN & BEINGN &  & \\
    
    \cline{1-3}
    DoS & & &\\
    DoS\,Hulk & & &\\ 
    DDoS & & &\\
    DoS\,GoldenEye & DoS/DDoS & 5.98 & ample\\ 
    DoS\,slowloris & & &\\
    DoS\,Slowhttptest & & &\\ 
    \cline{1-3}
    PortScan & PortScan & 14.31 & \\ 
    
    \hline
    FTP-Patator & Patator & & \\ 
    SSH-Patator &  &  164.33 & \\
    \cline{1-3}
    Web\,Attack-Brute\,Force &  & & \\ 
    Web\,Attack-XSS & Web\,Attack & 1042.28 & scarce \\ 
    Web\,Attack-Sql\,Injection &  & &\\ 
    
    \cline{1-3}
    Bot & Bot & 1156.92 & \\
    \hline
    Infiltration & Infiltration & 63141.03 & rare\\
    \cline{1-3}
    Heartbleed & Heartbleed & 206645.18 & \\
    \hline
\end{tabular}
    \label{tab:dataset}
\end{table}

By considering the step distribution of $I\!R_i$ values and its coherence with the visual depiction of the scatter, we classify intrusion detection traffic into ample-level, scarce-level, and rare-level (as shown in Table \ref{tab:dataset}), and treat them differently based on their respective attributes. 

To minimize the computational overhead while ensuring a high detection rate for the majority category, we specifically avoid processing the majority category (ample-level), which already exhibits a complete distribution.

In the scenario addressed in this paper, a challenge arises due to the significant disparity in the number of minority samples. Solely relying on data space-based data augmentation methods to generate minority samples may be ineffective for rare-level categories, as depicted in the lower right part of \cref{fig:scatter}, where the scarcity of samples hinders the generation of new instances. Conversely, employing only feature space-based data enhancement methods to synthesize minority samples may result in synthetic samples that closely resemble the original ones. Consequently, the performance of the scarce-level categories in the middle part of \cref{fig:scatter} may be constrained.

Consequently, we partition the minority categories into scarce-level and rare-level. For scarce-level categories, we adopt advanced data space-based data augmentation methods, while for rare-level categories, we rely on feature space-based data enhancement techniques.

The remaining parts of this paper are organized as follows. Section \ref{overview} gives the outline of our framework. Section \ref{design} introduces the main algorithms used to design the proposed S2CGAN. Section \ref{experiment} presents a detailed explanation of the architecture and results of the experiments conducted in this study.  Section \ref{related} reviews the literature on IDS and class imbalance problems. followed by Section \ref{conclusions}, which provides concluding remarks and suggests avenues for future research.

\section{Methodoloy Overview}
\label{overview}
\begin{figure}[!h]
    \centering
    \includegraphics[scale=0.7]{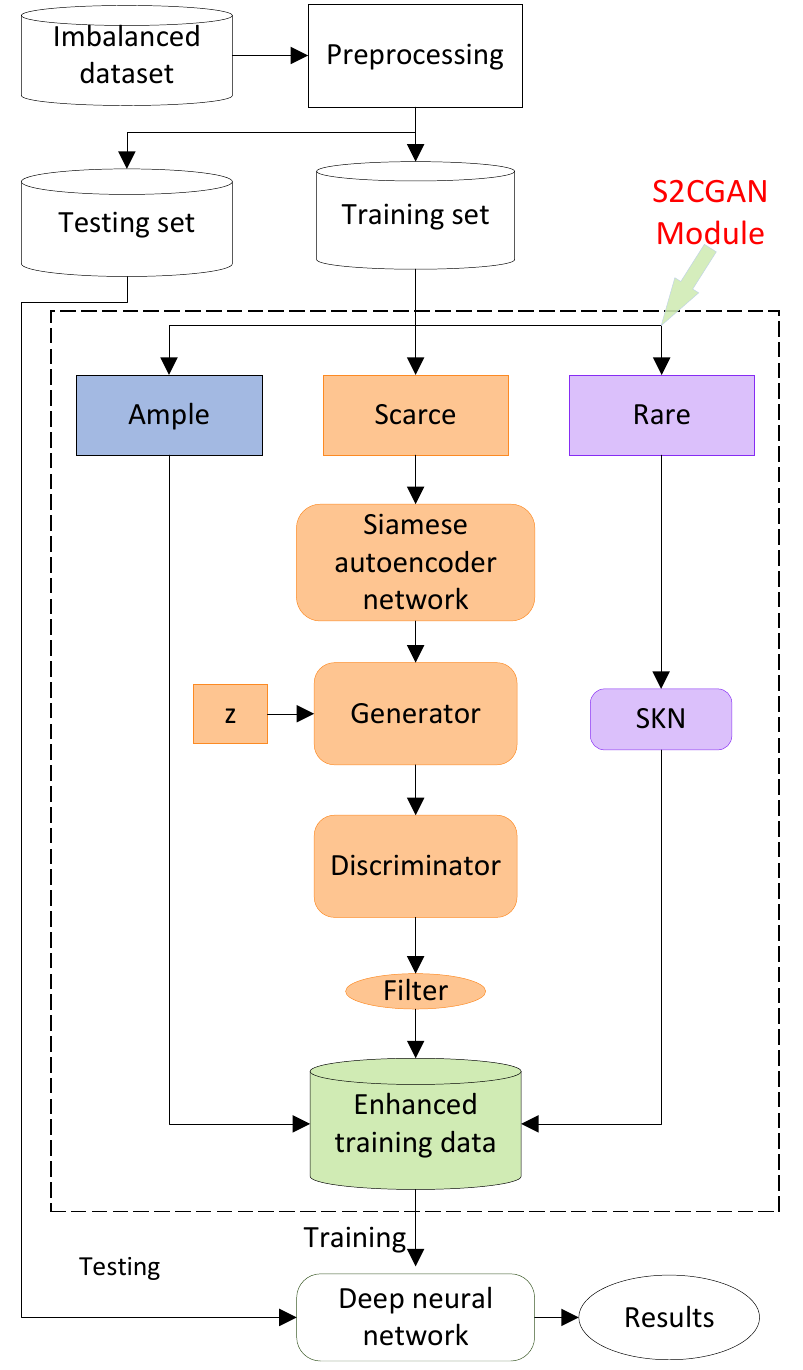}
    \caption{S2CGAN-IDS model framework. The framework diagram for the proposed algorithm consists of three primary components: data preprocessing, data enhancement, and training and testing of the IDS classifier.}
    \label{Figure:framework}
\end{figure}

In this section, we present our lightweight intrusion detection framework, which comprises three primary components as shown in \cref{Figure:framework}. 

These include dataset processing, the S2CGAN module, and classifier training and testing. Our framework is designed to improve the performance of intrusion detection systems in highly imbalanced datasets by employing different data augmentation techniques for different category levels. The specific detection process of the IDS framework is as follows.\\
\textbf{Dataset processing}. The dataset is processed through the following steps: normalization, and train-test split.\\
\textbf{S2CGAN module}. As a case study, we classify all categories within this dataset into three levels based on the step-change characteristics of their respective numbers. And employ the S2CGAN module to enhance the dataset, Which is a data generation model incorporating two techniques: SCGAN and SKN. The SCGAN is utilized to generate scarce-level attacks, while a filter is applied to the generated data to enhance the consistency of generated samples and original samples. On the other hand, SKN is used to generate rare-level attacks by simulating potential rare-level attack distributions through the KNN algorithm.\\
\textbf{Training/testing}. To evaluate the effectiveness of the proposed data augmentation algorithm, we implemented it using a deep neural network and trained an intrusion detection classifier using the augmented dataset. The performance of the resulting classifier was then verified using a separate test set.
\begin{algorithm}[!h]
  \SetAlgoLined
  \KwData{Original training dataset $T_o$\; 
  \qquad \quad  The SCGAN threshold $\eta$.}
  \KwResult{Argument training dataset $T_a$.}

  Initialize $T_{a}$ as an empty dataset\;
  Calculate $I\!R_i$ values of each class and divide them into ample-level, scarce-level, and rare-level based $I\!R_i$ rates from low to high\;
  \uIf{ample-level categories}{
    Add ample-level samples into $T_a$\;
  }
  \uElseIf{scarce-level categories}{
    Pretrain the SAE model by using Algorithm \ref{san}\;
    \For{each epoch} {
        Input the encoder result of SAE $s$ and random noise $z$ to CGAN\;
        Calculate the losses of the generator $G$ and the discriminator $D$\;
        Update the generator $G$ and the discriminator $D$ with their losses\;
    }
    Use the generator $G$ of well-trained SCGAN to generate new scarce-level samples $G(z|s)$\;
    \For{each $G(z|s)$}{
        Input $G(z|s)$ into the discriminator $D$ and output $(D(G(z|s))$\;
        \uIf{$(D(G(z|s)) \geq \eta)$}{
            Add $G(z|s)$ and its label into $T_a$\;
        }
    }
    Add scarce-level samples into $T_a$\;
  }
  \uElseIf{rare-level categories}{
    Synthesis new rare-level samples with Algorithm \ref{skn}\;
    Add the synthesized samples and original rare-level samples into $T_a$\;
  }
  \Return The augmented training dataset $T_a$\;
  \caption{S2CGAN (main)}
  \label{s2cgan}
\end{algorithm}

The S2CGAN module plays a central role in the algorithm proposed in this paper. This module enhances the original data set by operating in both the data and feature spaces, thus improving the detection accuracy of minority categories in the intrusion detection classifier. In the subsequent sections of this paper, we will present a detailed exposition of the S2CGAN (Algorithm \ref{s2cgan}).

\section{Implementation Details of S2CGAN}
\label{design}
As demonstrated in the principal component analysis (PCA) scatter plot of the CICIDS2017 dataset in \cref{fig:scatter}, the data distributions for ample-level categories are complete and can be directly reserved in the augmented dataset. However, scarce-level attacks have only approximate distribution. In this situation, we utilize SCGAN to learn the original distribution and generate missing data. Lastly, we employ SKN to expand the original distribution as much as possible from the feature space for rare-level attacks with only a few data points.

\subsection{SCGAN for scarce-level attacks}
This section is dedicated to the details of the SCGAN module, which serves the purpose of generating scarce-level categories. The SCGAN module consists of a Siamese autoencoder network (SAN) and a generative adversarial network (GAN). Firstly, we introduce the SAN model to extract differential feature information for the SCGAN module. Algorithm \ref{san} shows the detail.
\begin{algorithm}[!h]
    \SetAlgoLined
    \KwIn{Training dataset {$(x_1, y_1), (x_2, y_2), ..., (x_n, y_n)$}.}
    \KwOut{Model parameters $W_{s}$, $W_{d}$, $b_{s}$, $b_{d}$.}
    
    Initialize Encoder parameters $W_{s}$, $b_{s}$\;
    Initialize Decoder parameters $W_{d}$, $b_{d}$\;
    
    \For{each epoch} {
      Randomly select two different samples $(x_i, y_i)$ and $(x_j, y_j)$\;
      Feed each sample through a shared encoder to obtain their encoder outputs $f(x_i)$ and $f(x_j)$\;
      Use $f(x_i)$ and $f(x_j)$ as inputs to a shared decoder, producing the output $\bar{x}_i$ and $\bar{x}_j$\;
      Compute the loss function and backpropagate it through the network\;
      Update model parameters $W_{s}$, $W_{d}$, $b_{s}$, $b_{d}$\;}
    
    \Return model parameters $W_{s}$, $W_{d}$, $b_{s}$, $b_{d}$\;
    \caption{Siamese Autoencoder Networks (SAN)}
    \label{san}
\end{algorithm}

\begin{figure}[!h]
    \centering
    \includegraphics[scale=0.7]{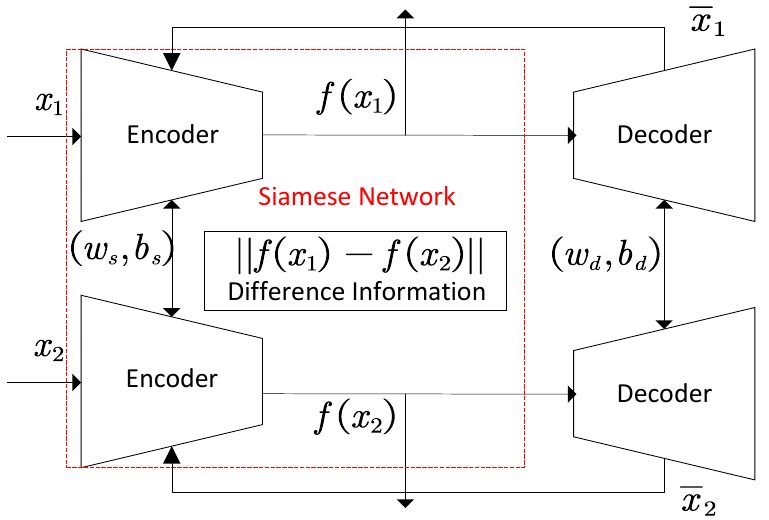}
    \caption{Siamese autoencoder networks. It consists of a pair of encoders sharing parameters and a pair of decoders sharing parameters.}
    \label{fig:san}
\end{figure}

SAN comprises a pair of autoencoders (AEs) with Siamese neural networks (SNNs). The SNNs are designed to capture the differences between the extracted key features of various attacks, while the AEs are responsible for extracting the most significant features. By integrating both models in the SAN, the features can be extracted more effectively and efficiently, resulting in the creation of scarce-level attacks. \cref{fig:san} provides a detailed illustration of the SAN.

SNN is a type of neural network that consists of two identical sub-networks, each taking a different input (i.e. $x_1$ and $x_2$) but with the same architecture, parameters, and weights. The two sub-networks output a pair of feature vectors $(f_w(x_1),f_w(x_2))$, which can be used to compute the similarity or difference between the two inputs. In this paper, the Euclidean distance (Eq. \eqref{Euler_dis}) was used to calculate the similarity between the two feature vectors. 
\begin{equation}
\label{Euler_dis}
    E_w (x_1,x_2)=||f_w(x_1)-f_w(x_2)||.
\end{equation}

The parameters can be optimized by minimizing the reconstruction error, which is typically computed with the following formula (Eq. \eqref{siamese_loss}):
\begin{equation}
\label{siamese_loss}
     \mathcal{L}_{S\!N\!N} = \frac{1}{2}\sum((1-y)E_w^2+y\max(0,m-E_w)^2),
\end{equation}
where $y$ is the binary label indicating whether the input pairs $(x_1,x_2)$ are similar ($y=0$) or dissimilar ($y=1$). $E_w^2$ in the loss function computes the squared distance between the feature vectors of similar pairs, while $\max(0,m-E_w)^2$ computes the squared distance between the feature vectors of dissimilar pairs, with a margin $m$ to ensure that the distance is smaller than $m$.

AE is a powerful unsupervised learning algorithm that is widely used in the field of machine learning. Unlike supervised learning methods, AE does not rely on labeled data for training. It consists of two critical components: encoder and decoder. The encoder is responsible for extracting the essential features $f_w(x)$ of the original data $x$, and the decoder aims to reconstruct the input data based on these extracted features. By minimizing the error between the reconstructed data $\bar{x}$ and the original input data during the learning process, AE learns the implicit feature representation of the data. In this study, the cost function is given by Eq.\eqref{autoencoder_loss}.
\begin{equation}
\label{autoencoder_loss}
     \mathcal{L}_{A\!E}(x,\bar{x})=\frac{1}{n}\sum_{i=1}^{n}(x_i - \bar{x}_i)^2.
\end{equation}

GAN is employed to learn the underlying data distribution and generate novel samples to fill in the missing data. It was first introduced by Goodfellow \cite{goodfellow2020generative} as a powerful approach for generating realistic images by training a generator network to produce images that are indistinguishable from real images by a discriminator network. Initially, GAN was used for generating realistic images from random noise, but they have since been extended to other domains, such as text and music generation. In addition to generating new data, GAN can also be used for other tasks, such as data augmentation, where the generator is used to produce additional training data for a given task. 

However, the original GAN generates samples that belong to a wide range of classes, without any control over the specific class of the generated samples. To overcome this limitation, conditional generative adversarial network (CGAN) \cite{mirza2014conditional} was introduced in the same year as an extension of GAN. CGAN incorporates conditional information into the generator G and the discriminator D outputs the probability of real and fake samples. This enables the generation of samples based on specific input conditions, allowing for targeted data synthesis.

The proposed SCGAN model aims to accelerate the convergence of CGAN. To achieve this goal, we employ difference information $s$ and random noise $z\in N(\mu,\sigma^2)$ obtained from the SAN model as input for CGAN. The generator $G$ takes $s$ and $z$ as the input and produces a set of pseudo samples $G(z|s)$, while the discriminator $D$ takes both the real attack samples $x$ and generated attack samples $G(z|s)$ as input.

The fundamental objective of SCGAN is demonstrated through $\mathop{min}\limits_{G}\mathop{max}\limits_{D} \mathcal{V}(D,G)$, as formula Eq. \eqref{V_GD}:
\begin{equation}
\label{V_GD}
     \mathcal{V}(D,G)=E_{x\sim{p_x}}log(D(x))+E_{z\sim{p_z}}log(1-D(G(z|s))).
\end{equation}

To train the discriminator $D$, we feed the conditional information $s$ and the noise vector $z$ into the generator $G$, which generates the pseudo samples $G(z|s)$. Subsequently, we feed both the real samples $x$ and the generated samples $G(z|s)$ into $D$ and update the parameters of $D$ based on the loss function $\mathcal{L}_D$ of the discriminator as shown in Eq. \eqref{L_D}.
\begin{equation}
\label{L_D}
    \mathcal{L}_D=-log(D(x))-log(1-D(G(z|s))).
\end{equation}

To train the generator $G$, we fix the parameters of the discriminator $D$ and the pseudo samples $G(z|s)$ into $D$. The error backpropagates to $G$ and its parameters are updated based on the loss function $\mathcal{L}_G$ to enhance its performance (Eq. \eqref{L_G}).
\begin{equation}
\label{L_G}
    \mathcal{L}_G=-log(D(G(z|s)).
\end{equation}

The alternating iterative training process of the SCGAN model persists until it reaches the stationary local Nash equilibrium \cite{heusel2018gans}, wherein the discriminator is unable to effectively differentiate between the pseudo samples and the real samples. At this point, both the generator and the discriminator have been effectively trained, enabling the SCGAN model to generate high-quality pseudo samples that exhibit a close resemblance to the actual attack samples.

After the generative adversarial network converges, the generated samples by the generator closely resemble real scarce-level attacks. However, it is possible for some noise points to still be present in the generated samples. To address this concern, a filter is applied after the discriminator to eliminate improperly generated samples. This ensures that the training of the intrusion detection classifier remains unaffected by the additional interference introduced by these samples.

\subsection{SKN for rare-level attacks} 
Rare-level attacks are differentiated from scarce-level attacks by their characteristics of having only a few scattered points. The sparsity of data at this level presents a challenge for deep neural networks to accurately simulate the possible distributions. To overcome this issue, we introduce a feature space-based oversampling method called SKN. This method specifically addresses the limited number of samples at this level, effectively avoiding the challenges associated with insufficient data.

Oversampling is a widely adopted technique to tackle imbalanced datasets, and popular methods include Random Over Sampling (ROS), SMOTE, and its variants. ROS involves replicating minority class instances to balance the class distribution while retaining the original information. However, it is susceptible to overfitting since it duplicates the same information multiple times. In contrast, SMOTE employs the KNN algorithm in the feature space to generate new samples and equalize the dataset. 
\begin{algorithm}[!h]
    \SetAlgoLined
    \KwIn{Training dataset $T_o$, where $X = {x_{1}, x_{2}, ..., x_{n}}$\;
          \qquad \quad \enspace Class distribution $C$, where $C = {c_{1}, c_{2}, ..., c_{l}}$\;
          \qquad \quad \enspace The number of class $i$ that need to be synthesized $N_{i}$\;
          \qquad \quad \enspace K-nearest neighbors parameter $k$.}
    \KwOut{Synthetically over-sampled dataset $T_{r}$.}

    Initialize $T_{r}$ as an empty dataset\;
    
    \For {each rare-level class $c_{i}$} {
      Find the k-nearest neighbors of each rare-level class sample $x_{i}$\;
      \For {each sample $x_{i}$} {
        Choose $N_i$ k-nearest neighbors randomly and denote them as $x_{1}, x_{2}, ..., x_{N_i}$\;
        \For {each $j=1,2,...,N_i$} {
            Choose a random number $\lambda\in[0, 1]$\;
            Set $x_{new} = x_{i} + \lambda \times (x_{j} - x_{i})$\;
            }
          Add the new sample $x_{new}$ and $c_i$ into $T_{r}$\;
          }
          }
\Return the over-sampled dataset $T_{r}$\;
    \caption{Synthesis K Neighbors(SKN)}
    \label{skn}
\end{algorithm}

In this study, we propose a novel rare-level oversampling method called SKN (Algorithm \ref{skn}), which leverages the advantages of existing algorithms to design a logically simple and efficient approach. The method involves three key steps. First, we identify $k$ adjacent samples of the $i\text{-}th$ rare class by means of a calculation process. Second, we randomly select $N_i$ samples from the adjacent set. Finally, we randomly apply a transformation function to the selected samples as well as the rare class samples themselves in order to synthesize new samples.

\section{Experiment}
\label{experiment}

\subsection{Benchmark dataset}
Our analysis of commonly used datasets from 1998 has revealed that most of them are outdated and unreliable. Some of these datasets suffer from limited traffic diversity and capacity, while others lack coverage of various known attacks. Additionally, some anonymize packet payload data, which makes them unable to reflect current trends. Moreover, several datasets lack essential features and metadata, which are necessary for accurate analysis.

For our research, we utilized the \href{https://www.unb.ca/cic/datasets/ids-2017.html}{CICIDS2017} dataset, which is a publicly available dataset composed of network traffic collected from the Canadian Institute for Cybersecurity (CIC) during weekdays in 2017. This dataset includes fifty sub-categories and seven major types of attacks and has gained popularity as a replacement for the previously widely used \href{https://kdd.ics.uci.edu/databases/kddcup99/kddcup99.html}{KDDCup99} and \href{https://www.unb.ca/cic/datasets/nsl.html}{NSLKDD} datasets due to its realistic and diverse traffic scenarios. The CICIDS2017 \cite{sharafaldin2018toward} dataset provides a reliable resource for studying network intrusion detection methods and has been widely adopted by researchers and practitioners in the field.

The CICIDS2017 dataset \cite{sharafaldin2018toward} consists of benign and recent common attacks. The dataset collection spanned five days and concluded on July 7, 2017, at 5:00 pm. The attacks were executed on Tuesday, Wednesday, Thursday, and Friday morning and afternoon, respectively. 

CICFlowMeter was utilized as a flow feature extraction tool to generate a CSV file with over 80 features based on the submitted PCAP file, which contains the flow data of the network interface card that can be obtained via Wireshark software or flow sniff function. There are two modes: online and offline. The online mode allows for real-time monitoring and feature generation, which can be saved locally after monitoring, whereas the offline mode entails the submission of a PCAP file and the receipt of a CSV file containing features.

After analyzing the dataset, we observed that some feature values exhibited significant variation. This wide disparity in feature values has the potential to result in slow network convergence and neuron output saturation. Therefore, we deemed it necessary to normalize the dataset. In this study, we employed the Min-Max normalization method, which resulted in the normalization of the data to a range of [0,1]. This approach enhances the comparability and compatibility of the features, mitigating the effects of the initial variation in the dataset and improving the overall quality of the results. The equation for the Min-Max normalization is as follows:
\begin{equation} \label{minmaxnor}
x_{norm}=\frac{x-x_{min}}{x_{max}-x_{min}},
\end{equation}
where $x_{norm}$ is the normalized value, $x$ is the original feature value, $x_{min}$ is the minimum value of the feature, and $x_{max}$ is the maximum value of the feature.

In accordance with best practices in machine learning, we partitioned the preprocessed dataset into training and test sets, with an 8:2 ratio respectively. It is important to note that the test set was only used during the evaluation phase of the IDS to ensure the validity and reliability of our experimental methodology.

We analyzed the imbalance ratio ($I\!R_i$) for all categories in the training set and observed a stepwise distribution of the imbalance ratio value. Notably, we identified a significant gap between the first and third echelons. To address this, we employed a categorization scheme where we grouped \emph{DoS/DDoS} and \emph{PortScan} attacks into the \textbf{ample} level, while \emph{Patator}, \emph{Web Attack}, and \emph{Bot} attacks were assigned to the \textbf{scarce} level. \emph{Infiltration} and \emph{Heartbleed} attacks were categorized as \textbf{rare} level. Table \ref{tab:dataset} shows the specific division of labels in the CICIDS2017 dataset and the corresponding attack levels according to the $I\!R_i$ team.

In this experiment, the CICIDS2017 dataset is randomly partitioned into training and test sets to validate the effectiveness of our approach. The training set is utilized to train the S2CGAN model and generate a sufficient number of samples as outlined in Table \ref{tab:train_test_split}. Subsequently, the IDS classifier is trained using the augmented dataset. To ensure unbiased and fair results, the test set is exclusively used to evaluate the performance of the IDS classifier.
\begin{table}[!h]
\centering
\caption{The detailed size of the training set, test set, and augmented datasets.}
\setlength{\tabcolsep}{1.5mm}
\begin{tabular}{ |c|c|c|c| }
    \hline
    \textbf{Category} & \textbf{Training(80\%)} & \textbf{Testing(20\%)} & \textbf{ After Augmentation} \\
    \hline
    BEINGN & 1818476 & 454620 & 1818476\\    
    \hline
    DoS/DDoS & 304550 & 76138 & 304550\\
    \hline
    PortScan & 127144 & 31785 & 127144\\     
    \hline
    Patator & 11068 & 2767 & 127144\\ 
    \hline
    Web Attack & 1744 & 436 & 127144\\  
    \hline
    Bot & 1573 & 393 & 127144\\  
    \hline
    Infiltration & 29 & 7 & 1573\\
    \hline
    Heartbleed & 9 & 2 & 1573\\
    \hline
\end{tabular}
    \label{tab:train_test_split}
\end{table}

\subsection{Evaluation metrics}
To comprehensively assess the performance of the proposed intrusion detection system in this paper, we employ Precision, Recall, and F-Score as evaluation metrics. These metrics are calculated at the sub-category level, using both weighted and macro averages. In this context, TP represents true positives, TN represents true negatives, FP represents false positives, and FN represents false negatives. These metrics provide a comprehensive evaluation of the system's performance in terms of accuracy, sensitivity, and overall effectiveness in detecting intrusions.

Precision is a metric that measures the proportion of correctly identified positive samples to all samples that were predicted as positive. Mathematically, it can be represented as:
\begin{equation}
    \label{pre}
    Precision=\frac{TP}{TP+FP}.
\end{equation}

Recall, also known as sensitivity or true positive rate, represents the number of samples of a specific class that were correctly identified out of all the samples belonging to that class. It can be calculated using the following equation:
\begin{equation}
    \label{recall}
    Recall=\frac{TP}{TP+FN}.
\end{equation}

The F-Score is a widely used metric that balances both Precision and Recall and provides a more comprehensive evaluation of the performance of a classifier. It is defined as the harmonic mean of Precision and Recall, and is given by the following formula:
\begin{equation}
\centering
    \label{f1}
    F_\beta=\frac{(1+\beta^2)Precision\times{Recall}}{\beta^2(Precision\times{Recall})}.
\end{equation}
The coefficient $\beta$ is used to describe the relative importance of Precision and Recall. For this particular experiment, we set $\beta$ as 1, which refers to the F1-score.

For the highly imbalanced dataset used in this experiment, using the weighted average formulas in Eq. \eqref{weighted_avg_eq} for the overall indicator may introduce a bias towards ample-level categories. This bias occurs due to the vast quantity of samples belonging to the ample-level categories compared to the minority categories. As a result, the weighted average may overly prioritize the performance of the ample-level categories and may not accurately reflect the performance of the minority-level categories.
\begin{equation}
\centering
\begin{split}
    \label{weighted_avg_eq}
    &P_{weighted}=\frac{\sum_{i=1}^n{w_i} \times {P}_i}{\sum_{i=1}^n{w_i}};\\
    &R_{weighted}=\frac{\sum_{i=1}^n{w_i} \times {R}_i}{\sum_{i=1}^n{w_i}};\\
    &{F_1}_{weighted}=\frac{\sum_{i=1}^n{w_i} \times {F_1}_i}{\sum_{i=1}^n{w_i}}.\\
\end{split}
\end{equation}
Here, $n$ is the total number of samples, $P_i$, $R_i$, and ${F_1}_i$ are the Precision, Recall, and F1-score values for class $i$, respectively, and $w_i$ is the weight assigned to class $i$, which is proportional to the frequency of that class in the dataset.
\begin{equation}
\centering
\begin{split}
    \label{macro_avg_eq}
    &P_{macro}=\frac{1}{n}\sum_{i=1}^n{P_i};\\
    &R_{macro}=\frac{1}{n}\sum_{i=1}^n{R_i};\\
    &{F_1}_{macro}=\frac{1}{n}\sum_{i=1}^n{{F_1}_i}.\\
\end{split}
\end{equation}

To evaluate the effectiveness of the S2CGAN model proposed in this paper for highly imbalanced data, the study primarily focuses on the macro average index. The macro average treats each category equally, making it a reliable measure of performance for attacks with a small number of categories. The formulas for calculating the macro average index are provided in Eq. \eqref{macro_avg_eq}. By giving equal weight to each category, the macro average provides a comprehensive assessment of the classifier's performance in handling imbalanced scenarios, particularly for rare-level categories.

\subsection{Experiment procedure}
This study utilizes the Keras and PyTorch frameworks to construct and evaluate the models. The experimentation is conducted on the Google Colaboratory Pro platform. The parameter settings and processing procedures for each module are described below.

The initial step in the preprocessing phase involves applying Min-Max normalization to the CICIDS2017 dataset, which restricts all numerical features to a range between 0 and 1. This normalization technique mitigates the influence of unit inconsistencies during the neural network training process. Subsequently, the dataset is randomly partitioned into a training set and a test set, with the training set representing 80\% of the total data.

Following the preprocessing phase, the training set is utilized to train the difference information extraction SAN module. The SAN module consists of two encoders and two decoders with shared parameters. The hidden layer of the SAN is configured as $64\times32\times16\times32\times64$, with the input and output layers comprising 78 neurons. The encoding dimension is set to 16, and the activation function employed is LeakyReLU. Each linear layer, excluding the output layer, is accompanied by BatchNorm1d, which enhances the connectivity within each batch. The batch size is specified as 64 to optimize the training process.

Following the categorization of the training set based on $I\!R_i$, the scarce-level attacks are passed through a pre-trained SAN to extract features that serve as the conditional information $s$ for the CGAN. In the CGAN architecture, the generator takes $s$ and Gaussian noise $z$ as inputs. The hidden layer parameters of the generator are configured as $32\times64\times128$, and the activation function employed for each layer, excluding the output layer, is LeakyReLU. The output layer utilizes the Sigmoid activation function. Each linear layer is accompanied by BatchNorm1d to facilitate batch-level connectivity.

The discriminator in the CGAN architecture consists of a hidden layer configured as $64\times8$. The activation function employed for each layer, excluding the output layer, is LeakyReLU, and the output layer utilizes the Sigmoid activation function. Each linear layer is followed by LayerNorm, which assists in normalizing the activations within each layer. The batch size for this module is set to 16 to optimize the training process.
\begin{figure}[!ht]
    \centering
    \includegraphics[scale=0.5]{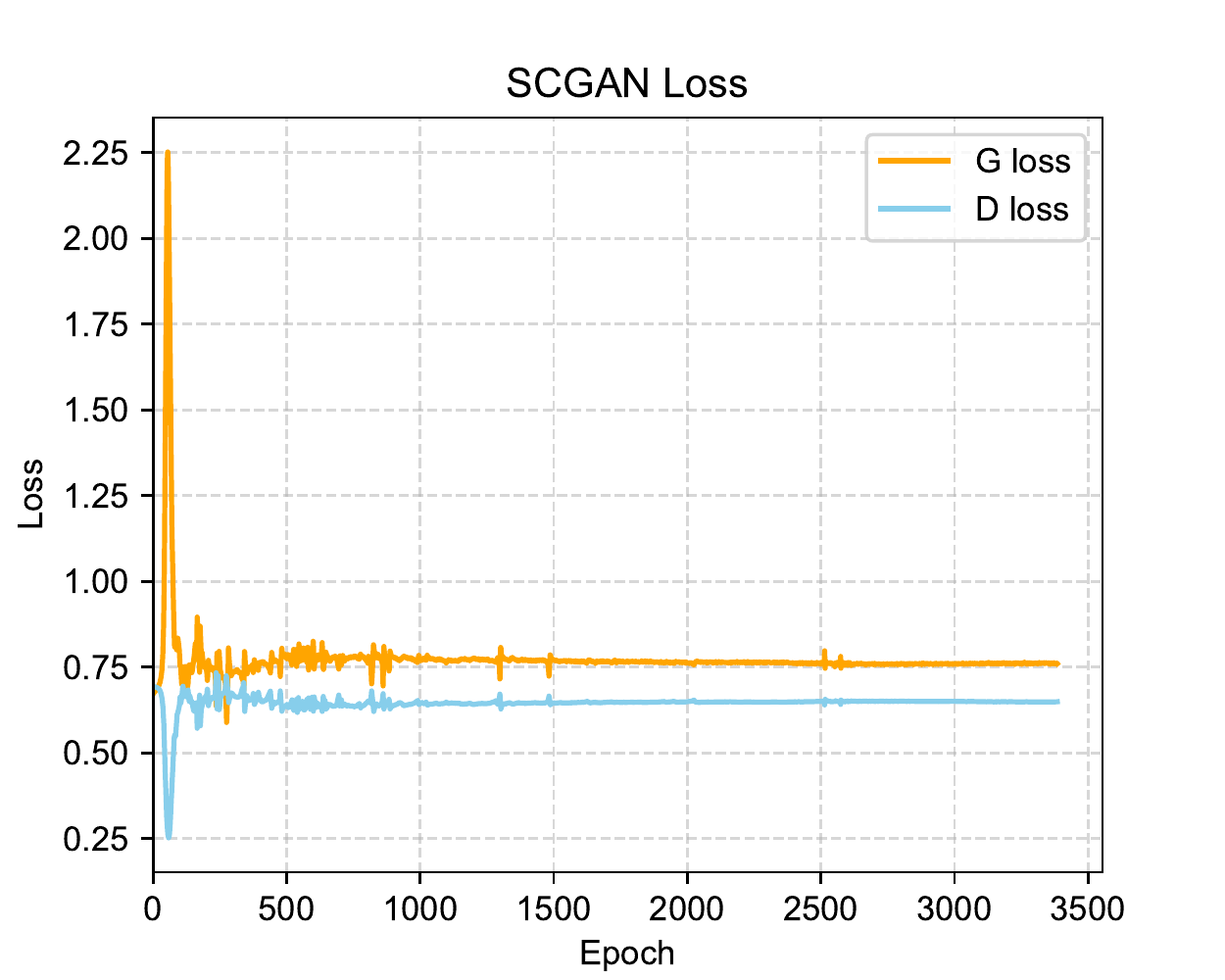}
    \caption{Losses of training SCGAN.}
    \label{fig:scgan_loss}
\end{figure}

Based on the experimental results as shown in \cref{fig:scgan_loss}, the loss functions of the generator and discriminator in the SCGAN model stabilize at a specific value after multiple rounds of adversarial training alternately them. This demonstrates that the fake samples generated by the generator can successfully deceive the discriminator, and it also shows that the SCGAN model almost convergence.

During the generation of scarce-level samples using the trained generator, we ensure sample quality by incorporating only those samples with a discriminator output exceeding 0.45 into the extended augmented dataset. For rare-level attacks, SKN is used to generate a sufficient number of samples (as Table \ref{tab:train_test_split}) to expand the augmented dataset.

\subsection{Comparative experiments}
\label{Comparative experiments}
This study entails a comparative analysis aimed at evaluating the performance of four algorithms for class imbalance. Our analysis encompasses the feature-based oversampling method SMOTE, as well as the data-based synthesis method CVAE-AN and the hybrid method TACGAN. The algorithms under investigation are as follows:\\
\textbf{Original}: This refers to the raw data without any processing.\\
\textbf{SMOTE}: Tesfahun et al. \cite{tesfahun2013intrusion} proposed SMOTE as a solution to address imbalanced datasets in machine learning. To generate a synthetic sample, SMOTE computes the difference between the feature vector of the considered instance and its nearest neighbor.\\
\textbf{CVAE-AN}: Sabeel et al. \cite{sabeel2021cvae} introduced a novel adversarial incremental learning approach called CVAE-AN. This approach employs a conditional variational autoencoder (CVAE) to generate new samples by learning a distribution from the training dataset. A Discriminator is then used to assess the quality of the generated samples based on how closely they resemble the original dataset.\\
\textbf{TACGAN}: Ding et al. \cite{ding2022imbalanced} introduced a tabular auxiliary classifier generative adversarial networks (TACGAN) model to address the issue of imbalanced intrusion detection systems. The proposed approach combines undersampling and oversampling techniques to tackle class imbalance. To be more precise, the majority class is undersampled using the KNN algorithm, while GAN is utilized to oversample a limited number of minority class samples.\\
\textbf{S2CGAN}: The method proposed in this paper.

\subsection{Discussion of experimental results}
To evaluate the effectiveness of S2CGAN for imbalanced network intrusion detection, this paper employs a basic deep neural network (DNN) as the intrusion detection classifier. The classifier consists of four hidden layers with $128\times64\times32\times16$, and the activation function for each layer is Relu.
\begin{table*}[!hbpt] 
\normalsize
\caption{Experimental results of multi-classification. Include the original classifier and four class imbalance algorithms.}
\label{tab:result_table}
\resizebox{\linewidth}{!}{
\begin{tabular}{|c|ccc|ccc|ccc|ccc|ccc|}
\hline
Methods & \multicolumn{3}{c|}{\textbf{Baseline}} & \multicolumn{3}{c|}{\textbf{CVAE+GAN}} & \multicolumn{3}{c|}{\textbf{SMOTE}} & \multicolumn{3}{c|}{\textbf{TACGAN}} & \multicolumn{3}{c|}{\textbf{S2CGAN}}\\
\hline
Metric & Precision & Recall & F1-Score & Precision & Recall & F1-Score & Precision & Recall & F1-Score & Precision & Recall & F1-Score & Precision & Recall & F1-Score \\
\hline
\small BEINGN & 0.9949 & \textbf{0.9916} & \textbf{0.9932} & \textbf{0.9979} & 0.9877 & 0.9928 & 0.9971 & 0.9847 & 0.9908 & 0.9965 & 0.9898 & 0.9931 & 0.9955 & 0.9903 & 0.9929 \\
\hline
\small DoS/DDoS & 0.9894 & \textbf{0.9994} & 0.9944 & \textbf{0.9904} & 0.9984 & 0.9944 & 0.9878 & 0.9993 & 0.9935 & \textbf{0.9904} & 0.9990 & \textbf{0.9947}	& 0.9896 & 0.9993 & 0.9944		 \\
\hline
\small PortScan & \textbf{0.9082} & 0.9369 & 0.9223 & 0.8671 & \textbf{0.9818} & 0.9209 & 0.8722 & 0.9644 & 0.9160 & 0.8878 & 0.9607 & \textbf{0.9228}	& 0.9016 & 0.9417 & 0.9212  \\
\hline
\small Patator & \textbf{0.9856} & 0.9884 & \textbf{0.9870} & 0.9682 & 0.9892 & 0.9785 & 0.9845 & 0.9895 & \textbf{0.9870}	& 0.9824 & 0.9895 & 0.9860 & 0.9433 & \textbf{0.9924} & 0.9672 \\
\hline
\small Web Attack & 0.9784 & 0.9358 & 0.9566 & 0.8881 & 0.8922 & 0.8902 & 0.2350 & 0.9404 & 0.3760 & 0.9000 & 0.9083 & 0.9041 & \textbf{0.9794} & \textbf{0.9794} & \textbf{0.9794}\\
\hline
\small Bot & \textbf{1.0000} & 0.3715 & 0.5417 & \textbf{1.0000} & 0.3740 & 0.5444 & 0.5746 & 0.7354 & 0.6451 & 0.9928 & 0.3486 & 0.5160 & 0.6091 & \textbf{0.7455} & \textbf{0.6705} \\
\hline
\small Infiltration & 0.0000 & 0.0000 & 0.0000 & 0.2500 & 0.1429 & 0.1818 & 0.8333 & \textbf{0.7143} & 0.7692 & 0.0000 & 0.0000 & 0.0000 & \textbf{1.0000} & \textbf{0.7143} & \textbf{0.8333} \\
\hline
\small Heartbleed & 0.0000 &  0.0000 &  0.0000 & 0.0000 & 0.0000 & 0.0000 & \textbf{1.0000} & \textbf{1.0000} & \textbf{1.0000} & 0.0000 & 0.0000 & 0.0000	& \textbf{1.0000} & \textbf{1.0000} & \textbf{1.0000} \\		
\hline
\end{tabular}
}
\end{table*}

Additionally, the output layer of the classifier is set to the number of attack categories, and the activation function is Softmax. The loss function used in the classifier is categorical cross-entropy, and the optimizer is Adam. The batch size is set to 128, and the maximum number of epochs is 100.

As previously discussed, intrusion detection systems often encounter highly imbalanced network traffic, where accurate identification of each type of attack is crucial for effective preventive measures and enhanced network security. The detailed findings of the study are presented in Table \ref{tab:result_table}, which provides a comprehensive overview of the results obtained from the evaluation of the proposed S2CGAN algorithm.

\begin{figure}[h]
    \centering
    \includegraphics[scale=0.4]{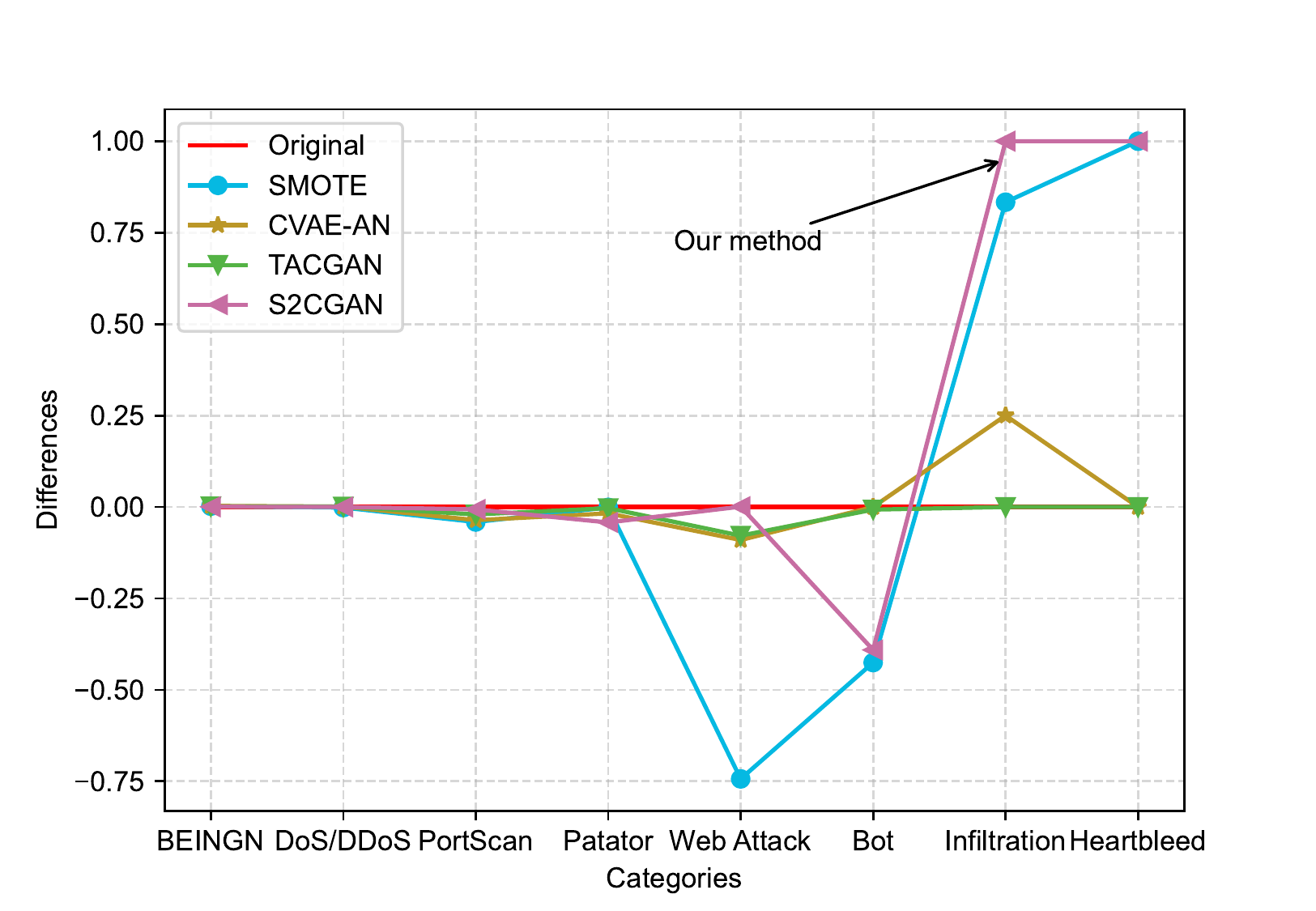}
    \caption{The difference in Precision between each method and the original dataset.}
    \label{fig:precision}
\end{figure}

\begin{figure}[h]
    \centering
    \includegraphics[scale=0.4]{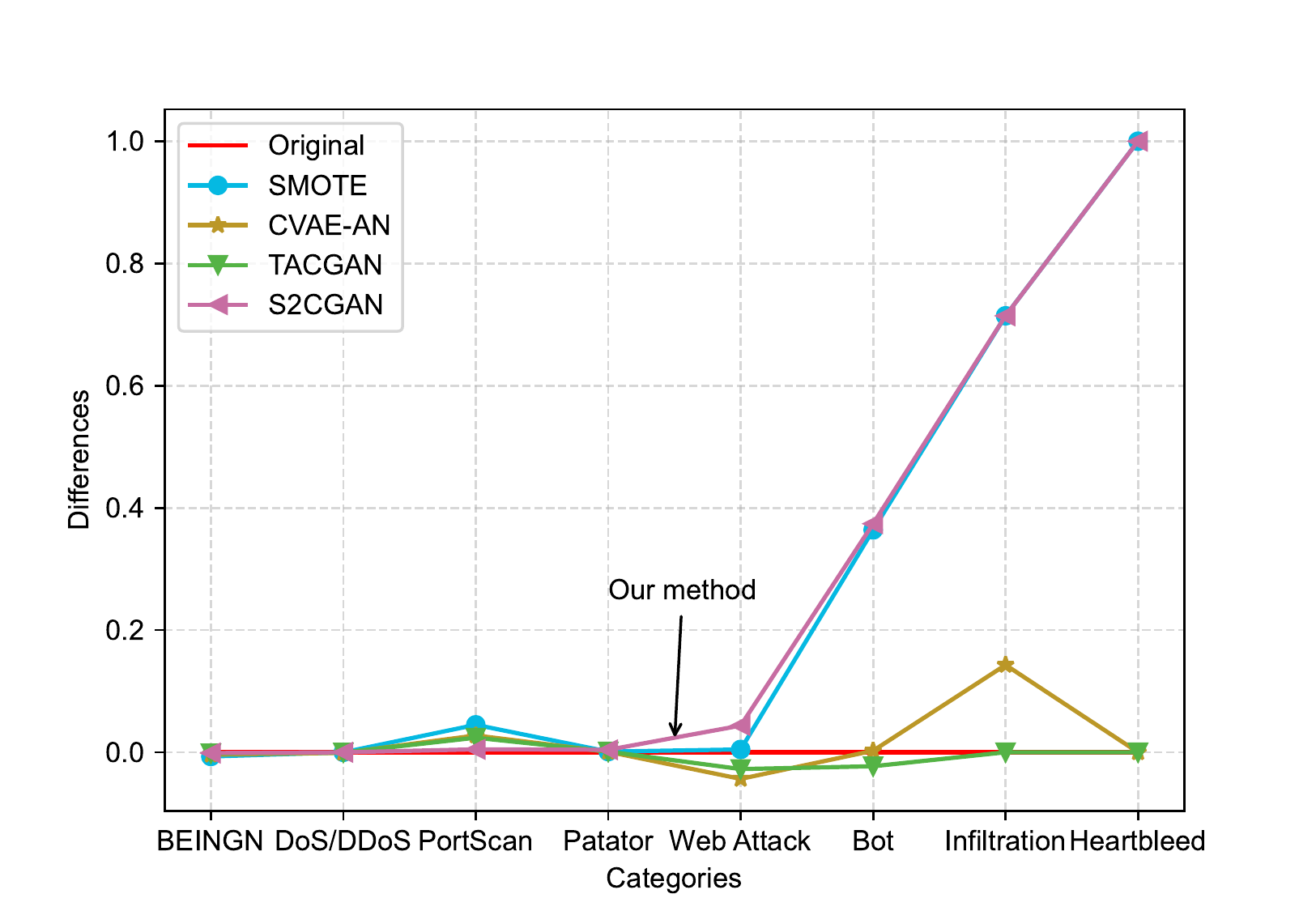}
    \caption{The difference in Recall between each method and the original dataset.}
    \label{fig:recall}
\end{figure}

\begin{figure}[h]
    \centering
    \includegraphics[scale=0.4]{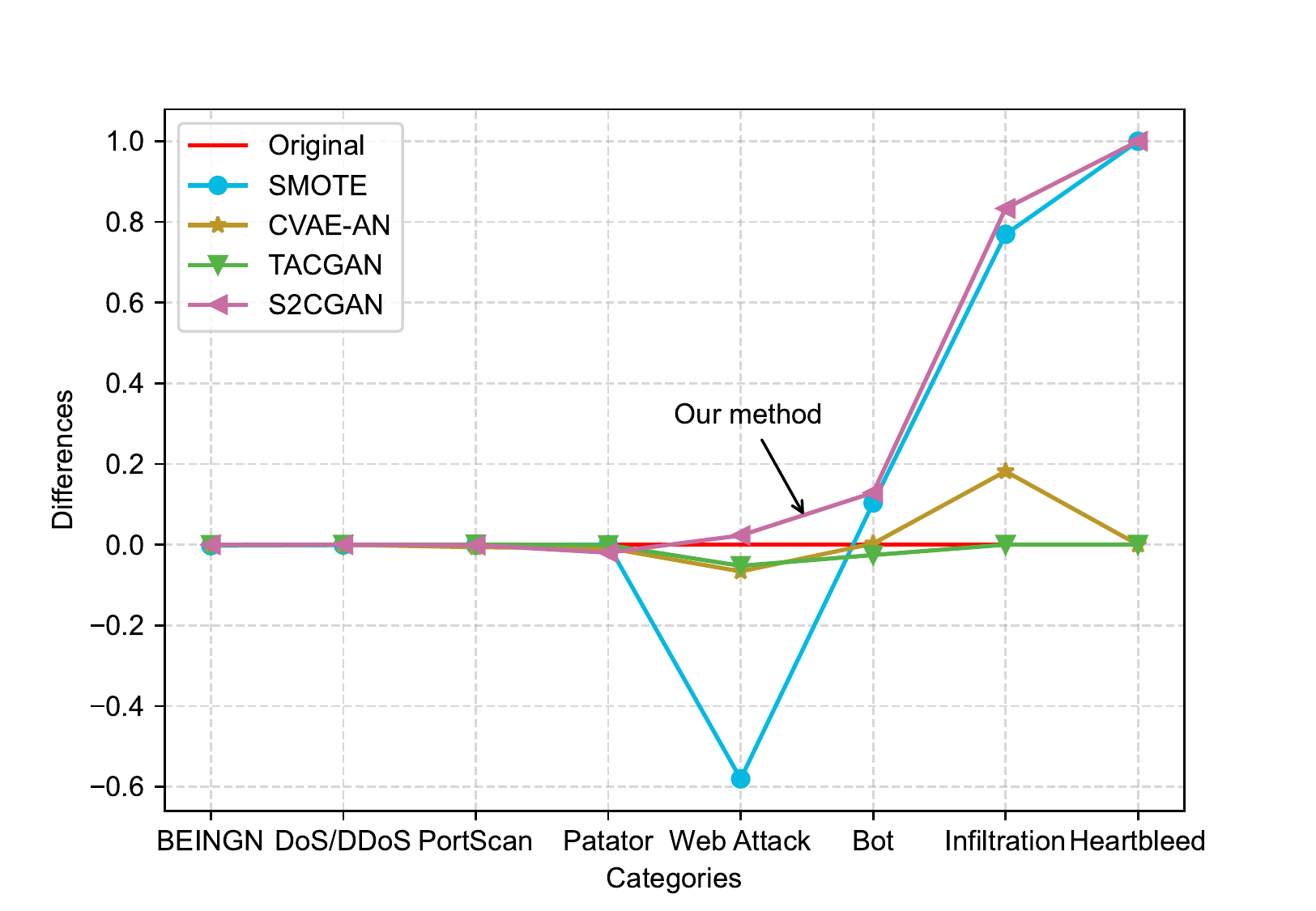}
    \caption{The difference in F1-score between each method and the original method.}
    \label{fig:f1}
\end{figure}

To facilitate a clear and comprehensive comparison of the performance of the four algorithms, we establish the IDS trained using the original dataset as the baseline. By analyzing and comparing the differences between the baseline and the four methods, we can gain insights into the effectiveness of each algorithm in improving the detection performance of the IDS. The details of our findings are presented in \cref{fig:precision}, \cref{fig:recall}, and \cref{fig:f1}, which depict the precision, recall, and F1-score, respectively. These figures provide a visual representation of the performance improvements achieved by each method across different attack categories, allowing for a comprehensive evaluation and comparison of their effectiveness in addressing the challenges posed by imbalanced network intrusion detection.

The performance analysis for the ample-level categories, as shown in the front sections of \cref{fig:precision} and \cref{fig:recall}, reveals that there is minimal variation in performance among the four methods employed in this study. Our method takes into consideration that ample-level samples exhibit a relatively comprehensive distribution following category classification. Therefore, no additional processing is conducted on samples of this level to minimize computational costs. Since the other three methods do not involve a separate class classification step, they process both ample-level samples and other samples in the same manner. Despite the additional computational cost incurred by SMOTE, CVAE-AN, and TACGAN, the expected performance improvements have not been observed at this level.
 
Regarding the scarce-level categories, the middle sections of \cref{fig:precision} and \cref{fig:recall} demonstrate that both our proposed method and SMOTE exhibit a slight decrease in Precision for the \emph{Patator} attack. Upon analyzing the scatter plot in \cref{fig:scatter}, it is evident that the distribution of the \emph{Patator} attack is highly concentrated. While our method successfully simulates the original distribution, it generates new points that lie outside the boundaries. However, it is reassuring that our method maintains a high Recall rate, ensuring the effective detection of the \emph{Patator} attack.

In the case of other scarce-level attacks, the SCGAN algorithm proposed in this study leverages the original data distribution to simulate the possible distribution of these scarce-level categories. This can be observed in the intermediate sections of \cref{fig:precision} and \cref{fig:recall}, where both Precision and Recall show improvements for the \emph{Web Attack} and \emph{Bot} categories. This indicates that our algorithm effectively captures and simulates the genuine data distribution of these two attack types, resulting in an enhanced detection rate for scarce-level attacks.

When considering rare-level attacks, the latter sections of \cref{fig:precision} and \cref{fig:recall} clearly show that both the baseline and TACGAN struggle to detect such attacks effectively. This is because the baseline and TACGAN rely solely on deep neural networks, which face difficulties in accurately representing the distribution of rare-level attacks due to the limited number of training samples available. On the other hand, CVAE-AN demonstrates a slight performance improvement as the variational encoder generates a larger number of slightly varied rare-level attacks. However, our method and SMOTE overcome the limitations of the data space by sampling from the eigenspace of rare-level attacks, taking into account the similarity of features among neighboring points in the eigenspace. The results clearly demonstrate that our method and SMOTE achieve outstanding performance in detecting rare-level attacks.

\cref{fig:f1} provides a visual representation of the difference in comprehensive F1-score between the four methods and the baseline for each category. The figure clearly demonstrates the significant performance advantage of our method over other approaches for all minority categories. Our method consistently achieves a higher F1-score, indicating its effectiveness in accurately detecting and classifying attacks across various minority categories.
\begin{figure}[!h]
    \centering
    \includegraphics[scale=0.4]{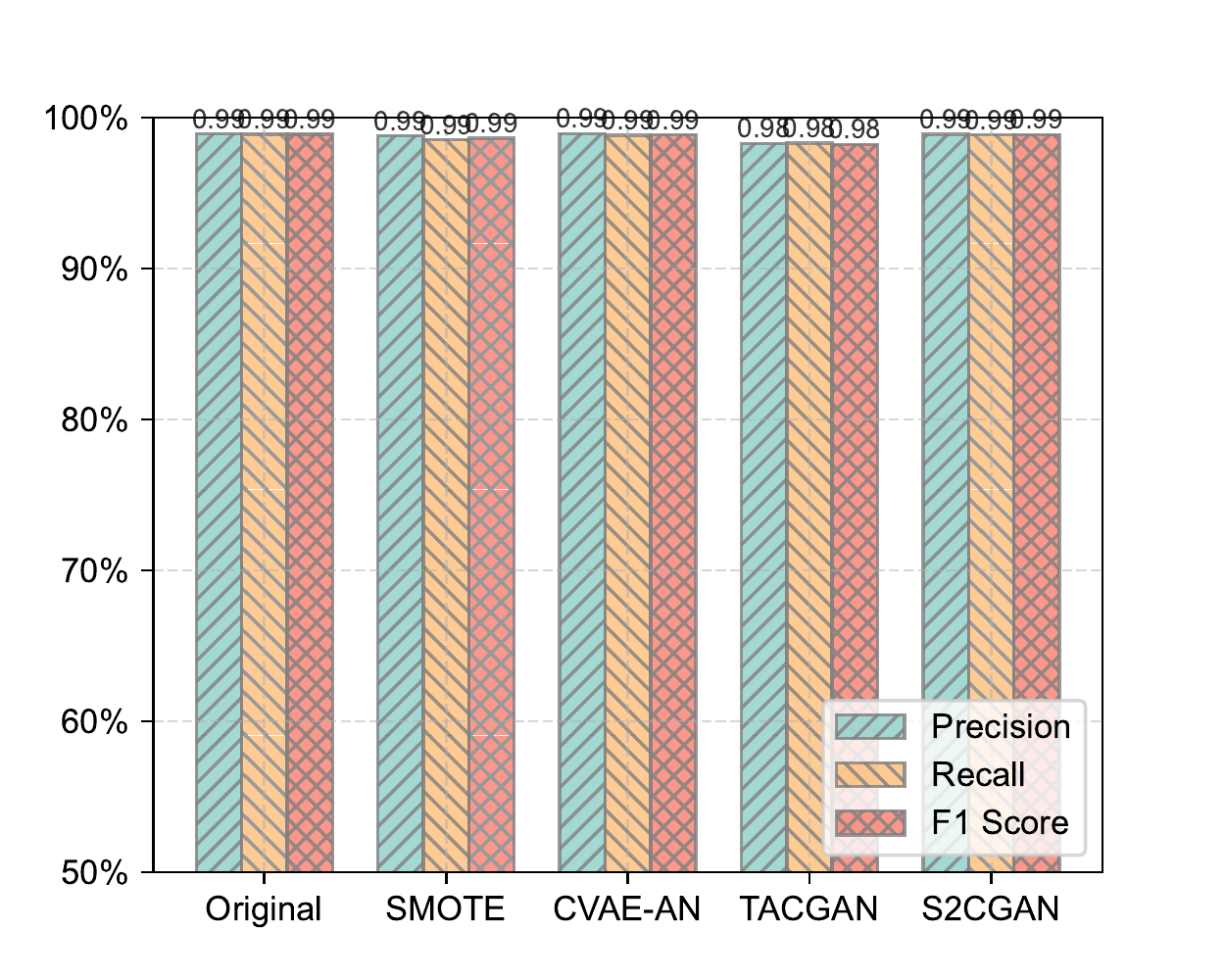}
    \caption{Weighted average: Each class is given different weights according to the frequency of each class.}
    \label{fig:weighted}
\end{figure}

\begin{figure}[!h]
    \centering
    \includegraphics[scale=0.4]{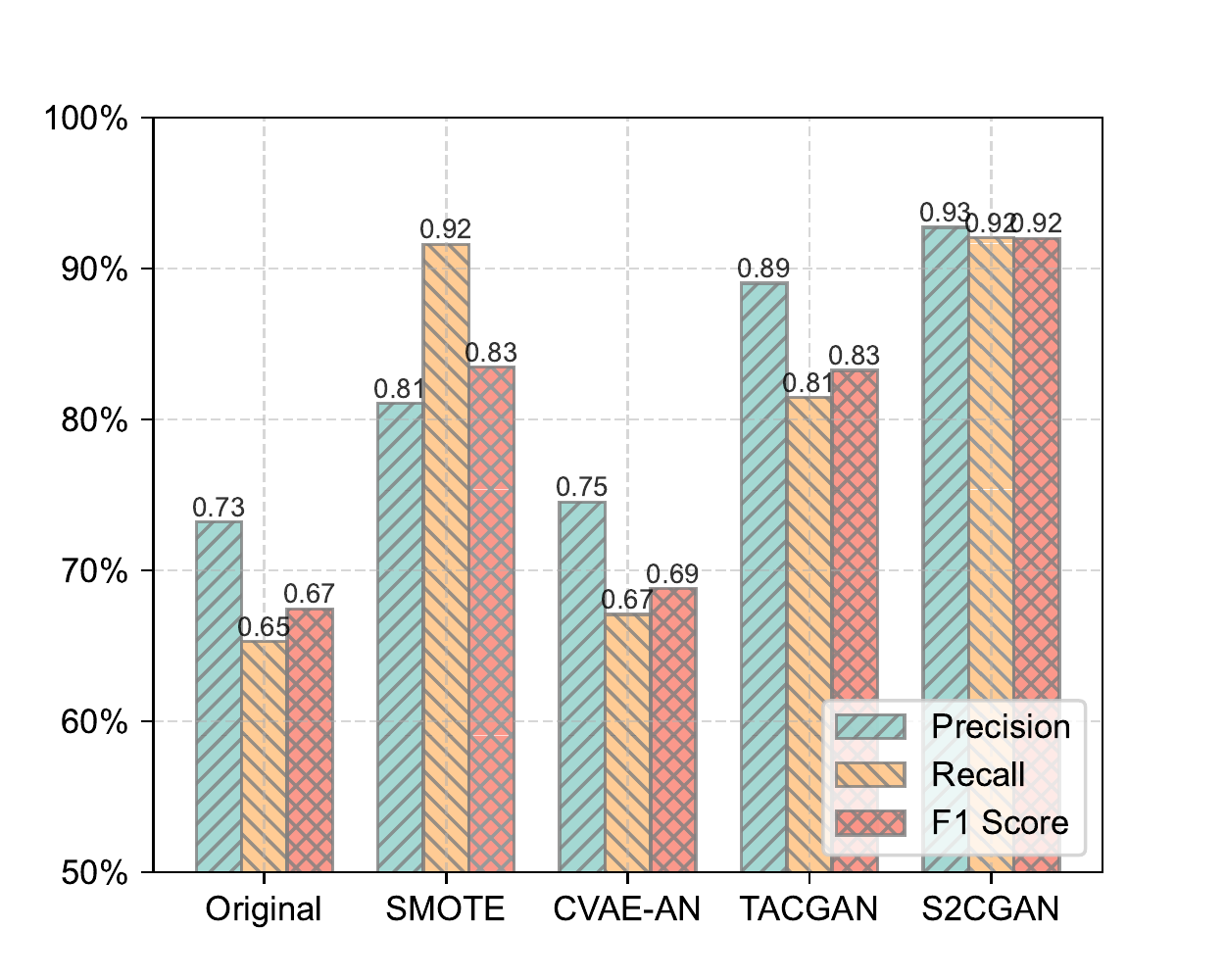}
    \caption{Macro average: Each class is given the same weight.}
    \label{fig:macro}
\end{figure}

When assessing the overall performance of the entire dataset, the weighted average is calculated by considering the frequency of occurrence of each category. This provides a comprehensive evaluation of the performance in unbalanced scenarios. \cref{fig:weighted} visually presents the comparison between our proposed framework and other methods across various categories. The figure indicates that our method performs similarly to other approaches regarding the weighted average, effectively ensuring a high detection rate for the overall dataset.

In IoT scenarios with less frequent attacks, using a single weighted average metric may not adequately capture the performance of minority categories. To overcome this limitation, we utilize the macro-average metric, which specifically evaluates the detection performance of classifiers in unbalanced scenarios. The macro average index assigns equal weight to each category, enabling a more accurate assessment of classifier performance in unbalanced settings. As depicted in \cref{fig:macro}, the macro-average index confirms that our framework successfully addresses the challenge of extreme class imbalance in intrusion detection systems. Our method demonstrates favorable outcomes even in highly imbalanced scenarios.

When comparing our method with the original classifier, significant improvements are observed in Precision, Recall, and F1-score. Specifically, our method demonstrates a remarkable increase of 22.05\% in Precision, a substantial improvement of 43.81\% in Recall, and a notable enhancement of 35.47\% in F1-score.

These results emphasize the effectiveness and superiority of our method in detecting attacks in highly imbalanced IoT networks. Our approach demonstrates exceptional performance in accurately identifying and classifying attacks, even in scenarios where the class distribution is heavily skewed. This highlights the practical relevance and robustness of our method in improving the security and reliability of IoT networks.

\section{Related works}
\label{related}
Intrusion detection systems are an essential part of IoT network security, playing a crucial role in detecting malicious network activities. With the proliferation of IoT devices and the increasing sophistication of attacks, there is a growing need for robust and efficient IDS. In recent years, deep learning has emerged as a promising approach to enhance the effectiveness of IDS. In this section, we provide a concise overview of the literature on deep learning-based intrusion detection, with a specific focus on addressing the problem of class imbalance in deep intrusion detection systems.

\subsection{Deep learning-based IDS}
The utilization of deep learning has gained substantial popularity in the domain of intrusion detection.  Various deep learning techniques, including deep belief networks (DBNs) \cite{wei2019optimization, suzen2021developing}, convolutional neural networks (CNNs) \cite{kim2020cnn,mendoncca2021intrusion,nguyen2020genetic}, and recurrent neural networks (RNNs) \cite{yin2017deep,almiani2020deep}, have been employed for this purpose. 

Deep learning has demonstrated its efficacy in detecting abnormal patterns in network traffic and identifying potential intrusions. However, in order to achieve optimal performance, deep learning models necessitate comprehensive data distribution. Regrettably, intrusion detection systems often encounter the challenge of class imbalance, leading to inadequate performance of deep learning models, particularly in detecting attacks that belong to the minority class.

\subsection{Class imbalance in IDS}
The issue of class imbalance is prevalent in intrusion detection, where the number of normal samples outweighs the number of attack samples. To tackle this challenge, researchers have proposed various techniques from three different perspectives: the loss function or structure of the classifier, the synthesis of minority class samples from the feature space, and the generation of new minority class samples from the data space.

In terms of the classifier, Wang et al. \cite{wang2016training} proposed a new loss function called mean false error (MFE) along with its improved version, mean squared false error (MSFE), that captures errors of both majority and minority classes equally, providing a solution to the data imbalance problem in deep networks from an evaluation perspective. Bedi et al. \cite{bedi2020siam, bedi2021siamids} employed Siamese neural networks in Siam-IDS and its improved variant, I-SiamIDS, to achieve higher recall values for both R2L and U2R attack classes. Meanwhile, Gupta et al. \cite{gupta2021lio, gupta2022cse} proposed two intrusion detection methods: CSE-IDS, which integrates the eXtreme gradient boosting (XGBoost) algorithm with sensitive unbalanced deep intrusion detection, and LIO-IDS, based on long short-term memory (LSTM) and One-vs-One algorithms, both of which demonstrate high detection rates and reduced computational costs.

However, these approaches often come at the expense of sacrificing the detection accuracy of majority categories, resulting in an increased detection rate for minority categories. This trade-off undermines the reliability of intrusion detection systems.

For the method of synthesizing minority classes from the feature space, Al et al. \cite{al2021stl} developed the CNN-LSTM method, which combines a hybrid deep learning (HDL) approach with STL (SMOTE + Tomek-Link) class imbalance processing to enhance intrusion detection performance. Hasib et al. \cite{hasib2021hsdlm} proposed a hybrid method that combines KNN undersampling with SMOTE oversampling to enhance the dataset of network intrusion detection systems. 

Merely synthesizing new samples of minority categories from the feature space alone is insufficient to capture the comprehensive information present in the high-dimensional space of minority category attacks. Consequently, the detection rate of these methods for minority categories remains severely limited.

Fortunately, the emergence of GAN has brought about new techniques for generating new samples of minority classes from the data space. These advancements bring renewed hope for effectively addressing the class imbalance problem in IDS. 

Lee et al. \cite{lee2021gan} proposed the use of GAN to generate synthetic data to balance datasets and improve the performance of imbalanced IDS. Recent studies have shown that GANs can effectively generate realistic attack traffic \cite{ali2019mfc} and enhance the accuracy of intrusion detection models \cite{zhou2020deep}. Huang \cite{huang2020igan} proposed Imbalance Generative Adversarial Networks (IGAN) to generate representative samples for minority classes, while Cui et al. \cite{cui2023novel} introduced a Wasserstein GAN (WGAN) module to address unbalanced data. 

These approaches have shown notable advancements in addressing the class imbalance between minority and majority classes compared to generating synthetic data solely from the feature spaces. However, in the domain of wireless networks, there are highly rare attacks that can inflict significant damage on network devices if they occur. Regrettably, the aforementioned method of generating new attacks from the data space is not effective in addressing these types of highly rare attacks. In such scenarios, alternative approaches need to be explored to tackle the challenges posed by these exceptionally rare attack instances. 

\section{Conclusions and Future Work}
\label{conclusions}
Intrusion detection is a critical technology that is essential for ensuring IoT network security. However, the problem of extreme class imbalance can severely compromise the performance of intrusion detection systems. To address this issue, we propose a lightweight framework for intrusion detection called S2CGAN-IDS. Our framework includes a data processing module that categorizes network traffic into three levels: ample, scarce, and rare, based on the degree of imbalance. We then employ an efficient SCGAN model to generate new scarce-level attacks and use simple SKN to oversample rare-level attacks. Finally, we train a simple DNN classifier with the augmented dataset. Our experimental findings demonstrate the superiority of our method over other approaches for addressing the class imbalance, as evidenced by the macro average metric. Notably, our framework exhibits enhanced detection rates for nearly all minority classes while maintaining a high detection rate for the majority class.

In the future, we plan to explore additional classification models to further validate the effectiveness of our proposed scheme. Additionally, we acknowledge that this paper is a case study in division category levels, and we aim to apply our S2CGAN-IDS to more general scenes in future research. By doing so, we hope to gain a deeper understanding of the generalizability and scalability of our proposed approach, and its potential to improve intrusion detection performance on a broader range of datasets.

\section{Acknowledgement}
\label{Acknowledgement}
This work was partially supported by the National Natural Science Foundation of China (62171085).


\ifCLASSOPTIONcaptionsoff
  \newpage
\fi

\bibliographystyle{IEEEtran}
\bibliography{IEEEabrv,S2CGAN_IDS}

\begin{thebibliography}{10}
\providecommand{\url}[1]{#1}
\csname url@samestyle\endcsname
\providecommand{\newblock}{\relax}
\providecommand{\bibinfo}[2]{#2}
\providecommand{\BIBentrySTDinterwordspacing}{\spaceskip=0pt\relax}
\providecommand{\BIBentryALTinterwordstretchfactor}{4}
\providecommand{\BIBentryALTinterwordspacing}{\spaceskip=\fontdimen2\font plus
\BIBentryALTinterwordstretchfactor\fontdimen3\font minus
  \fontdimen4\font\relax}
\providecommand{\BIBforeignlanguage}[2]{{%
\expandafter\ifx\csname l@#1\endcsname\relax
\typeout{** WARNING: IEEEtran.bst: No hyphenation pattern has been}%
\typeout{** loaded for the language `#1'. Using the pattern for}%
\typeout{** the default language instead.}%
\else
\language=\csname l@#1\endcsname
\fi
#2}}
\providecommand{\BIBdecl}{\relax}
\BIBdecl

\bibitem{liao2013intrusion}
H.-J. Liao, C.-H.~R. Lin, Y.-C. Lin, and K.-Y. Tung, ``Intrusion detection
  system: A comprehensive review,'' \emph{Journal of Network and Computer
  Applications}, vol.~36, no.~1, pp. 16--24, 2013.

\bibitem{mohamed2011acoustic}
A.-r. Mohamed, G.~E. Dahl, and G.~Hinton, ``Acoustic modeling using deep belief
  networks,'' \emph{IEEE transactions on audio, speech, and language
  processing}, vol.~20, no.~1, pp. 14--22, 2011.

\bibitem{girshick2015fast}
R.~Girshick, ``Fast r-cnn,'' in \emph{Proceedings of the IEEE international
  conference on computer vision}, 2015, pp. 1440--1448.

\bibitem{schuster1997bidirectional}
M.~Schuster and K.~K. Paliwal, ``Bidirectional recurrent neural networks,''
  \emph{IEEE transactions on Signal Processing}, vol.~45, no.~11, pp.
  2673--2681, 1997.

\bibitem{lecun2015deep}
Y.~LeCun, Y.~Bengio, and G.~Hinton, ``Deep learning,'' \emph{nature}, vol. 521,
  no. 7553, pp. 436--444, 2015.

\bibitem{aminanto2016deep}
E.~Aminanto and K.~Kim, ``Deep learning in intrusion detection system: An
  overview,'' in \emph{2016 International Research Conference on Engineering
  and Technology (2016 IRCET)}.\hskip 1em plus 0.5em minus 0.4em\relax Higher
  Education Forum, 2016.

\bibitem{liu2019machine}
H.~Liu and B.~Lang, ``Machine learning and deep learning methods for intrusion
  detection systems: A survey,'' \emph{applied sciences}, vol.~9, no.~20, p.
  4396, 2019.

\bibitem{ahmad2021network}
Z.~Ahmad, A.~Shahid~Khan, C.~Wai~Shiang, J.~Abdullah, and F.~Ahmad, ``Network
  intrusion detection system: A systematic study of machine learning and deep
  learning approaches,'' \emph{Transactions on Emerging Telecommunications
  Technologies}, vol.~32, no.~1, p. e4150, 2021.

\bibitem{liu2022intrusion}
C.~Liu, R.~Antypenko, I.~Sushko, and O.~Zakharchenko, ``Intrusion detection
  system after data augmentation schemes based on the vae and cvae,''
  \emph{IEEE Transactions on Reliability}, vol.~71, no.~2, pp. 1000--1010,
  2022.

\bibitem{zhang2019deep}
H.~Zhang, X.~Yu, P.~Ren, C.~Luo, and G.~Min, ``Deep adversarial learning in
  intrusion detection: A data augmentation enhanced framework,'' \emph{arXiv
  preprint arXiv:1901.07949}, 2019.

\bibitem{khan2017cost}
S.~H. Khan, M.~Hayat, M.~Bennamoun, F.~A. Sohel, and R.~Togneri,
  ``Cost-sensitive learning of deep feature representations from imbalanced
  data,'' \emph{IEEE transactions on neural networks and learning systems},
  vol.~29, no.~8, pp. 3573--3587, 2017.

\bibitem{zhang2018cost}
C.~Zhang, K.~C. Tan, H.~Li, and G.~S. Hong, ``A cost-sensitive deep belief
  network for imbalanced classification,'' \emph{IEEE transactions on neural
  networks and learning systems}, vol.~30, no.~1, pp. 109--122, 2018.

\bibitem{dhote2020hybrid}
S.~Dhote, C.~Vichoray, R.~Pais, S.~Baskar, and P.~Mohamed~Shakeel, ``Hybrid
  geometric sampling and adaboost based deep learning approach for data
  imbalance in e-commerce,'' \emph{Electronic Commerce Research}, vol.~20, pp.
  259--274, 2020.

\bibitem{aljawarneh2018anomaly}
S.~Aljawarneh, M.~Aldwairi, and M.~B. Yassein, ``Anomaly-based intrusion
  detection system through feature selection analysis and building hybrid
  efficient model,'' \emph{Journal of Computational Science}, vol.~25, pp.
  152--160, 2018.

\bibitem{wu2019transfer}
P.~Wu, H.~Guo, and R.~Buckland, ``A transfer learning approach for network
  intrusion detection,'' in \emph{2019 IEEE 4th international conference on big
  data analytics (ICBDA)}.\hskip 1em plus 0.5em minus 0.4em\relax IEEE, 2019,
  pp. 281--285.

\bibitem{singla2019overcoming}
A.~Singla, E.~Bertino, and D.~Verma, ``Overcoming the lack of labeled data:
  Training intrusion detection models using transfer learning,'' in \emph{2019
  IEEE International Conference on Smart Computing (SMARTCOMP)}.\hskip 1em plus
  0.5em minus 0.4em\relax IEEE, 2019, pp. 69--74.

\bibitem{zhang2020novel}
J.~Zhang, M.~Zheng, J.~Nan, H.~Hu, and N.~Yu, ``A novel evaluation metric for
  deep learning-based side channel analysis and its extended application to
  imbalanced data,'' \emph{IACR Transactions on Cryptographic Hardware and
  Embedded Systems}, pp. 73--96, 2020.

\bibitem{le2022xgboost}
T.-T.-H. Le, Y.~E. Oktian, and H.~Kim, ``Xgboost for imbalanced multiclass
  classification-based industrial internet of things intrusion detection
  systems,'' \emph{Sustainability}, vol.~14, no.~14, p. 8707, 2022.

\bibitem{leevy2021mitigating}
J.~L. Leevy, T.~M. Khoshgoftaar, and J.~M. Peterson, ``Mitigating class
  imbalance for iot network intrusion detection: a survey,'' in \emph{2021 IEEE
  Seventh International Conference on Big Data Computing Service and
  Applications (BigDataService)}.\hskip 1em plus 0.5em minus 0.4em\relax IEEE,
  2021, pp. 143--148.

\bibitem{stiawan2022improvement}
D.~Stiawan, M.~Y.~B. Idris, S.~Defit, Y.~S. Triana, R.~Budiarto \emph{et~al.},
  ``Improvement of attack detection performance on the internet of things with
  pso-search and random forest,'' \emph{Journal of Computational Science},
  vol.~64, p. 101833, 2022.

\bibitem{goodfellow2020generative}
I.~Goodfellow, J.~Pouget-Abadie, M.~Mirza, B.~Xu, D.~Warde-Farley, S.~Ozair,
  A.~Courville, and Y.~Bengio, ``Generative adversarial networks,''
  \emph{Communications of the ACM}, vol.~63, no.~11, pp. 139--144, 2020.

\bibitem{mirza2014conditional}
M.~Mirza and S.~Osindero, ``Conditional generative adversarial nets,''
  \emph{arXiv preprint arXiv:1411.1784}, 2014.

\bibitem{heusel2018gans}
M.~Heusel, H.~Ramsauer, T.~Unterthiner, B.~Nessler, and S.~Hochreiter, ``Gans
  trained by a two time-scale update rule converge to a local nash
  equilibrium,'' 2018.

\bibitem{sharafaldin2018toward}
I.~Sharafaldin, A.~H. Lashkari, and A.~A. Ghorbani, ``Toward generating a new
  intrusion detection dataset and intrusion traffic characterization.''
  \emph{ICISSp}, vol.~1, pp. 108--116, 2018.

\bibitem{tesfahun2013intrusion}
A.~Tesfahun and D.~L. Bhaskari, ``Intrusion detection using random forests
  classifier with smote and feature reduction,'' in \emph{2013 International
  Conference on Cloud \& Ubiquitous Computing \& Emerging Technologies}.\hskip
  1em plus 0.5em minus 0.4em\relax IEEE, 2013, pp. 127--132.

\bibitem{sabeel2021cvae}
U.~Sabeel, S.~S. Heydari, K.~Elgazzar, and K.~El-Khatib, ``Cvae-an: Atypical
  attack flow detection using incremental adversarial learning,'' in \emph{2021
  IEEE Global Communications Conference (GLOBECOM)}.\hskip 1em plus 0.5em minus
  0.4em\relax IEEE, 2021, pp. 1--6.

\bibitem{ding2022imbalanced}
H.~Ding, L.~Chen, L.~Dong, Z.~Fu, and X.~Cui, ``Imbalanced data classification:
  A knn and generative adversarial networks-based hybrid approach for intrusion
  detection,'' \emph{Future Generation Computer Systems}, vol. 131, pp.
  240--254, 2022.

\bibitem{wei2019optimization}
P.~Wei, Y.~Li, Z.~Zhang, T.~Hu, Z.~Li, and D.~Liu, ``An optimization method for
  intrusion detection classification model based on deep belief network,''
  \emph{Ieee Access}, vol.~7, pp. 87\,593--87\,605, 2019.

\bibitem{suzen2021developing}
A.~A. S{\"u}zen, ``Developing a multi-level intrusion detection system using
  hybrid-dbn,'' \emph{Journal of Ambient Intelligence and Humanized Computing},
  vol.~12, no.~2, pp. 1913--1923, 2021.

\bibitem{kim2020cnn}
J.~Kim, J.~Kim, H.~Kim, M.~Shim, and E.~Choi, ``Cnn-based network intrusion
  detection against denial-of-service attacks,'' \emph{Electronics}, vol.~9,
  no.~6, p. 916, 2020.

\bibitem{mendoncca2021intrusion}
R.~V. Mendon{\c{c}}a, A.~A. Teodoro, R.~L. Rosa, M.~Saadi, D.~C. Melgarejo,
  P.~H. Nardelli, and D.~Z. Rodr{\'\i}guez, ``Intrusion detection system based
  on fast hierarchical deep convolutional neural network,'' \emph{IEEE Access},
  vol.~9, pp. 61\,024--61\,034, 2021.

\bibitem{nguyen2020genetic}
M.~T. Nguyen and K.~Kim, ``Genetic convolutional neural network for intrusion
  detection systems,'' \emph{Future Generation Computer Systems}, vol. 113, pp.
  418--427, 2020.

\bibitem{yin2017deep}
C.~Yin, Y.~Zhu, J.~Fei, and X.~He, ``A deep learning approach for intrusion
  detection using recurrent neural networks,'' \emph{Ieee Access}, vol.~5, pp.
  21\,954--21\,961, 2017.

\bibitem{almiani2020deep}
M.~Almiani, A.~AbuGhazleh, A.~Al-Rahayfeh, S.~Atiewi, and A.~Razaque, ``Deep
  recurrent neural network for iot intrusion detection system,''
  \emph{Simulation Modelling Practice and Theory}, vol. 101, p. 102031, 2020.

\bibitem{wang2016training}
S.~Wang, W.~Liu, J.~Wu, L.~Cao, Q.~Meng, and P.~J. Kennedy, ``Training deep
  neural networks on imbalanced data sets,'' in \emph{2016 international joint
  conference on neural networks (IJCNN)}.\hskip 1em plus 0.5em minus
  0.4em\relax IEEE, 2016, pp. 4368--4374.

\bibitem{bedi2020siam}
P.~Bedi, N.~Gupta, and V.~Jindal, ``Siam-ids: Handling class imbalance problem
  in intrusion detection systems using siamese neural network,'' \emph{Procedia
  Computer Science}, vol. 171, pp. 780--789, 2020.

\bibitem{bedi2021siamids}
------, ``I-siamids: an improved siam-ids for handling class imbalance in
  network-based intrusion detection systems,'' \emph{Applied Intelligence},
  vol.~51, pp. 1133--1151, 2021.

\bibitem{gupta2021lio}
N.~Gupta, V.~Jindal, and P.~Bedi, ``Lio-ids: Handling class imbalance using
  lstm and improved one-vs-one technique in intrusion detection system,''
  \emph{Computer Networks}, vol. 192, p. 108076, 2021.

\bibitem{gupta2022cse}
------, ``Cse-ids: Using cost-sensitive deep learning and ensemble algorithms
  to handle class imbalance in network-based intrusion detection systems,''
  \emph{Computers \& Security}, vol. 112, p. 102499, 2022.

\bibitem{al2021stl}
S.~Al and M.~Dener, ``Stl-hdl: A new hybrid network intrusion detection system
  for imbalanced dataset on big data environment,'' \emph{Computers \&
  Security}, vol. 110, p. 102435, 2021.

\bibitem{hasib2021hsdlm}
K.~M. Hasib, N.~A. Towhid, and M.~R. Islam, ``Hsdlm: a hybrid sampling with
  deep learning method for imbalanced data classification,''
  \emph{International Journal of Cloud Applications and Computing (IJCAC)},
  vol.~11, no.~4, pp. 1--13, 2021.

\bibitem{lee2021gan}
J.~Lee and K.~Park, ``Gan-based imbalanced data intrusion detection system,''
  \emph{Personal and Ubiquitous Computing}, vol.~25, pp. 121--128, 2021.

\bibitem{ali2019mfc}
A.~Ali-Gombe and E.~Elyan, ``Mfc-gan: Class-imbalanced dataset classification
  using multiple fake class generative adversarial network,''
  \emph{Neurocomputing}, vol. 361, pp. 212--221, 2019.

\bibitem{zhou2020deep}
F.~Zhou, S.~Yang, H.~Fujita, D.~Chen, and C.~Wen, ``Deep learning fault
  diagnosis method based on global optimization gan for unbalanced data,''
  \emph{Knowledge-Based Systems}, vol. 187, p. 104837, 2020.

\bibitem{huang2020igan}
S.~Huang and K.~Lei, ``Igan-ids: An imbalanced generative adversarial network
  towards intrusion detection system in ad-hoc networks,'' \emph{Ad Hoc
  Networks}, vol. 105, p. 102177, 2020.

\bibitem{cui2023novel}
J.~Cui, L.~Zong, J.~Xie, and M.~Tang, ``A novel multi-module integrated
  intrusion detection system for high-dimensional imbalanced data,''
  \emph{Applied Intelligence}, vol.~53, no.~1, pp. 272--288, 2023.

\end{thebibliography}

\balance





\end{document}